\documentclass[12pt]{iopart}

\begin{document}

\title[Spacetime could be simultaneously discrete and continuous]{Spacetime
could be simultaneously continuous
and discrete in the same way that information can}

\author{Achim Kempf}

\address{Departments of Applied Mathematics and Physics, University of Waterloo \\ 200
University Avenue West, Waterloo N2T 3G1, Ontario, Canada}
\ead{akempf@uwaterloo.ca}
\begin{abstract}
There are competing schools of thought about the question of whether spacetime
is fundamentally either continuous or discrete. Here, we consider the possibility that
spacetime could be simultaneously continuous \it and \rm discrete, in the same mathematical way that
information can be simultaneously continuous and discrete. The equivalence of continuous and discrete information, which is of key importance in signal processing, is established by Shannon sampling theory: of any bandlimited signal it suffices to record discrete samples to be able to perfectly reconstruct it everywhere, if the samples are taken at a rate of at least twice the bandlimit. It is known that
physical fields on generic curved spaces obey a sampling theorem if they possess an ultraviolet cutoff. Most recently, methods of spectral geometry have been employed to show that also the very shape of a curved space (i.e., of a Riemannian manifold) can be discretely sampled and then reconstructed up to the cutoff scale. Here, we develop these results further, and we here also consider the generalization to curved spacetimes, i.e., to Lorentzian manifolds.

\end{abstract}

\maketitle

\section{Introduction}
A well-known gedanken experiment on the basis of general relativity and the
uncertainty principle indicates that the notion of distance loses operational
meaning at scales below the Planck length of $10^{-35}m$: assume that distances
could be resolved with an uncertainty $\Delta x$ that is smaller than the Planck length. The resulting momentum
uncertainties $\Delta p$ imply, via the Einstein equation, that
there are curvature uncertainties. These then would be large enough to prevent distances from being knowable with an uncertainty as small as $\Delta x$.

Due to the continuing lack of experiments that can reach the Planck scale, it is still not known how the structure of spacetime at the Planck scale is to be described in the much sought-after theory of quantum gravity. It is clear only that from some larger length scale, $l_{QFT}$, onwards the fundamental
quantum gravity theory must reduce to quantum field theory (QFT) on
curved continuous spacetimes. In fact, the success of inflationary cosmology indicates
that the length scale $l_{QFT}$ at which QFT on curved space emerges from quantum gravity
is likely at most about five orders of magnitude larger than the Planck
length. This is because successful predictions of inflationary cosmology hinge on the
assumption that QFT holds on curved spacetime for length scales as small as the
Hubble radius during inflation, which evidence indicates
was likely only around five or six orders of magnitude larger than the Planck length.

The closeness of the Planck scale and the scale
of the Hubble radius during inflation makes it conceivable that future
precision tests of inflation, e.g., by measuring the B-polarisation spectrum of
the cosmic microwave background, could eventually provide experimental
data against which to test theoretical models of spacetime at the Planck scale. To obtain predictions, it is of great interest, therefore, to explore the range of theoretical possibilities for the short-distance structure of spacetime.

\section{Candidate spacetime structures}

There are two basic theoretical possibilities. One is that
spacetime is ultimately correctly described as a continuous manifold, as in string theory, and the
other would be that spacetime is fundamentally discrete. See, e.g., \cite{discreteqg}. But are there other possibilities for the structure of spacetime?

The first possibility that we will now briefly consider (for completeness) is likely not
realistic. But the example is perhaps instructive nevertheless, since it shows how quickly the above question leads one to the deepest levels of mathematics. Namely, let us consider the possibility that
spacetime possesses an intermediate short-distance structure, in between
continuous and discrete. Concretely, we ask whether perhaps spacetime could ultimately be mathematically described as a set
of points (or events) whose cardinality is in between countable infinity and
continuous infinity. This question leads indeed to the very heart of mathematics,
namely the incompleteness results of logic:
Recall that Cantor had hypothesized that there cannot exist any set of a cardinality between countable and continuous infinity. G\"odel and Cohen \cite{Goedel,Cohen} then showed that Cantor's continuum hypothesis (CH) is a concrete example that demonstrates the fundamental incompleteness of mathematical theories: CH is neither provable nor disprovable within standard (i.e., ZF) set theory. Relevant for physical theories is the implication that it is therefore impossible to construct a set of intermediate cardinality in any explicit
way. This is because else CH could be disproved, which would contradict Cohen and G\"odel's
finding that CH is independent of the axioms of standard set theory. This leads us to disregard the possibility that spacetime may have a cardinality between discrete and continuous. In this
context see, however, also \cite{Isham}.

How else, then, could one account for the phenomenon predicted by the gendanken experiment above, namely the phenomenon that the very notion of distance should lose operational meaning at around the Planck scale? One possibility is that spacetime is discrete, in which case there are indeed operationally no physical distances smaller than the lattice spacing. Another possibility is that spacetime could still be a continuum - and that
the notion of distance loses operational meaning beyond a cutoff length scale because all particles are fundamentally non-pointlike, i.e., extended objects such as strings or membranes.

Much is known, of course, about both these possibilities, from large bodies of work on discrete quantum gravity theories on one hand and string theory on the other. For example, it is known that in theories that assume spacetime to be discrete it can be difficult
to obtain a continuous spacetime manifold of fixed dimension in the low energy limit. In
theories that describe spacetime as continuous, such as string theory,
$0$-dimensional dynamical objects, such as $0$-branes can re-emerge.
Here, we will consider a further possibility which has recently been suggested \cite{ak-prl-2009},
namely the possibility that spacetime could be simultaneously continuous and discrete, in the
same mathematical way that information can be simultaneously continuous and discrete.

\section{Sampling theory}

 The starting point here is Shannon's
sampling theorem, \cite{shannon}, which is at the heart of information theory.
The sampling theorem provides the crucial equivalence between continuous and
discrete representations of information and it is of ubiquitous use in
communication engineering and signal processing. Concretely, the theorem states
that any function (the signal), $f(t)$, whose frequency spectrum is bounded
by some finite value, $\Omega$, (the bandlimit) can be reconstructed \it
perfectly \rm for all $t$ from the amplitude samples $f(t_n)$ (e.g., raw music
data file) taken at equidistant sample points $t_n$ if their spacing $\delta t:= t_{n+1}-t_n$
is at most $(2\Omega)^{-1}$:
\begin{equation}
f(t) = \sum_n \mbox{sinc}(2(t-t_n)\Omega)~f(t_n)\label{shannon-o}
\end{equation}
Note that the accuracy of the reconstruction of the continuous signal $f$
is limited only by the accuracy with which the samples $f(t_n)$ were taken. There is a vast body of literature on sampling theory, much of it addressing practical matters such as the effects of lost samples or time and amplitude measurement inaccuracies, see e.g., \cite{shannon}.
The generalization to multiple dimensions, via the cartesian product,
is straightforward and allows one to apply sampling theory, for example, to images. The bandlimit of an image, e.g., an image made via a telescope, is of course determined by the aperture. Similarly, sampling theory can be used straightforwardly to reconstruct any bandlimited physical field $\phi$ in $3$-dimensional flat space from samples $\phi(x_n)$ taken at a sufficiently dense set of points $\{x_n\}$. This was first pointed out in \cite{ak-prl-2000}, where it was shown that this case is implied by a minimum length uncertainty principle that arose from general studies of quantum gravity and from string theory.

Regarding the application of sampling theory in physics, the key questions then are:
how can sampling theory be generalized to fields on curved spaces, and on
curved spacetimes? And, can one also develop a sampling theory for curved
spaces and curved spacetimes themselves? The latter should be a sampling theory
that allows one, from the knowledge of suitable samples taken at discrete points of a curved manifold,
to reconstruct the very ``shape'' of the manifold,
except for structure (or ``wrinkles") that are smaller than the cutoff scale.
Manifolds that differ only by wrinkles that are smaller than the
cutoff scale would not be physically distinguishable and should presumably
represent the same spacetime. Our aim now is to collect the partial answers that have been reached so far and then to extend the theory.

Let us begin by considering the underlying principle that allows
one in certain circumstances to do the seemingly impossible, namely to
perfectly reconstruct a continuous function from samples of its amplitudes taken at a discrete set of points. To
this end, consider first the case of an $N$-dimensional function space, $F$,
spanned by some generic basis functions $\{b_i(x)\}_{i=1...N}$. Then, all
functions $f\in F$ obey $f(x)=\sum_{i=1}^N\lambda_i ~b_i(x)$ for some
$\{\lambda_i\}$. For the functions in any such finite-dimensional function space,
$F$, there automatically holds a sampling theorem: assume that of a
function $f\in F$ only its amplitudes $a_n=f(x_n)$ are known for $n=1...N$ at
some $N$ generically-chosen points $x_n$. Thus, we have:
\begin{equation}
f(x_n)=a_n=\sum_{i=1}^N \lambda_i ~b_i(x_n)\label{e1}
\end{equation}
Then, Eq.\ref{e1} generally allows one to determine the coefficients $\lambda_i$
and therefore $f(x)$ for all $x$. This is because for generic basis functions,
$\{b_i\}$, and generic sample points $\{x_n\}$, the determinant of the $N\times N$
matrix, $B$, defined through $B_{n,i}:=b_i(x_n)$ for $i,n=1...N$ is non-vanishing and the matrix is therefore
invertible. We obtain $\lambda_i=\sum_{j=1}^N B^{-1}_{ij} ~a_j$ and therefore:
\begin{equation}
f(x) = \sum_{n=1}^N f(x_n) G(x_n,x) \mbox{~~~~for all}~x
\end{equation}
Here, the so-called reconstruction kernel, $G$, reads: $G(x_n,x)=\sum_{i=1}^N
B^{-1}_{ni}~b_i(x)$. Sampling theorems for infinite-dimensional function
spaces, such as Shannon's in Eq.\ref{shannon-o}, rely on the same principle, though proving them using this method is more subtle. First, we truncate the function space to finite dimensions,
where a sampling theorem holds by this general principle, and we then carefully take the limit in which the truncation is removed.

Of course, instead of using the general principle in this way, one could easily prove Eq.\ref{shannon-o} using Fourier theory, as textbooks on sampling theory usually do. However, Fourier theory will not be general enough for our purposes. For example, Fourier theory does not naturally generalize to non-equidistantly-taken samples and to curved spaces.

Let us collect important features of sampling theory. From the general principle it is clear that
samples need not be taken equidistantly because all that is required is that the determinant of $B$ is nonzero. Indeed, also for the case of the infinite-dimensional space of $\Omega$-bandlimited functions it is known that perfect reconstruction from non-equidistant samples is possible - as long as the sample points' average density, technically the Beurling density, is at least $2\Omega$. The general principle also indicates that the reconstruction may become numerically instable, however, e.g., when the determinant of $B$ becomes very small. Indeed, in the case of the infinite dimensional space of bandlimited functions,
the reconstruction of $f$ from samples $f(t_n)$ becomes numerically less and less stable as the samples are taken more and more non-equidistantly. The reconstruction can be shown to be most stable, i.e., least sensitive to numerical errors in the sample values, when the samples are taken equidistantly. Of interest will also be the fact that whenever sampling theory applies, there is a strict equivalence of integration and summation: for example for equidistantly-sampled $\Omega$-bandlimited functions, one has: $\sum_n\phi(t_n)\psi(t_n)=2\Omega\int dt~\phi(t)\psi(t)$. This fact has been used, for example, to evaluate hard-to-sum series ({\it e.g.}, in
analytic number theory): view the terms of the series in question as samples of
bandlimited functions, rewrite the series as an integral and then apply
powerful integration tools such as integration by parts or contour integration (that would otherwise not have counterparts in the theory of series).
Analogous sampling theoretic tools may be useful in quantum gravity, for example, to turn series into analytically easier-to-handle integrals, or to turn integrals into numerically easier-to-handle series.

\section{Overview: sampling theory in physical theories}

The idea that physical fields could possess the sampling property, i.e., that
they could be reconstructed everywhere from merely discretely taken samples,
was first proposed in \cite{ak-prl-2000}. There, it was also shown that the fields of any theory
naturally acquire the sampling property if, in the language of first
quantization, the uncertainty relations are modified in the ultraviolet so that
there is formally a finite lower bound, $\Delta x_{min}$, on the uncertainty in
position. Such uncertainty relations have indeed arisen in various studies of
quantum gravity and string theory, see {\it e.g.}, \cite{ucrs,ak-qheisenberg}.

In the simplest case, a physical theory on flat space contains a bandwidth-type natural UV-cutoff, i.e., the fields do not contain wavelengths shorter than say the Planck length. The theory's fields, equations of motion and actions, which live on a continuous space, are then determined everywhere from their samples on a sufficiently dense lattice.  In fact, as in Shannon sampling, the theory can be formulated on any lattice whose average spacing is tight enough (i.e., at the cutoff
length scale). Since no sampling lattice is preferred,
external symmetries, such as possible Killing vector fields need not be broken by the discretization.
This of course also follows directly from the fact that the theory is, via sampling theory, equivalent to a theory whose fields live on a continuum.

This finding is of interest also in the context of work on the possible effects of quantum gravity on spacetime symmetries.  For example, with the mathematical discovery of quantum groups, it has long been suggested that Planck scale effects could slightly deform external and or internal Lie group symmetries in field theories into quantum group symmetries. In fact, generalized uncertainty relations that arose from quantum group symmetric Heisenberg algebras led to the first examples of sampling-theoretic cutoffs, \cite{ak-qheisenberg}.  Also, for example, in \cite{Gambini} it was shown that Planck scale physics could affect the propagation of light by making the vacuum a medium with helicity-dependent dispersion. This would lead to potentially observable effects whose scale could be bounded, for example, by precision measurements of gamma ray bursts.  If spacetime exhibits a bandlimitation at the Planck scale, then this too can lead to possibly observable effects, e.g.,  in the cosmic microwave background's B-polarization, as we will discuss below, \cite{cosmo}. It has also been suggested \cite{ak-lattice}, that the degrees of freedom that are being cut off  by the bandlimitation may re-emerge as the internal degrees of freedom of gauge theories. The idea is that the gauge principle expresses the nonobservability of spatial structure below the cutoff scale. This idea is supported by the observation that, on general functional analytic grounds, the family of optimal lattice discretizations in sampling theory is necessarily parametrized  by unitary groups, \cite{ak-lattice}. Further, a key question is of course whether quantum gravity effects break local Lorentz invariance. In \cite{Amelino}, it was pointed out that and how an observer-independent preferred length scale can be consistent with local Lorentz invariance. Also the sampling-theoretic cutoff can be covariant, e.g., if the spectrum of the d'Alembertian is cut off, since the d'Alembertian is a scalar operator. In the spacetime covariant case there is no shortest wavelength, of course. However, as was shown in \cite{ak-prl-2004}, modes of wavelengths that are much shorter than the Planck length are dynamically effectively frozen since their temporal bandwidth is exceedingly small. The Lorentz contraction of a modes' spatial wavelength and the time dilatation of a mode's temporal bandwidth together make this phenomenon covariant. Clearly, more research into the possible effects of quantum gravity on local external and internal symmetries is needed. 

Notice that in our discussion of sampling theory so far we have discussed mostly the case of flat space, disregarding time, and that we have not been concerned with spacetime covariance yet. As the next step, let us now consider the introduction of curvature, i.e., covariant sampling theory on curved spaces (not spacetimes). This includes the case of Wick-rotated spacetimes and, for example, also spacelike hypersurfaces in cosmology.

In this case, the UV cutoff, i.e., the ``bandlimitation" for
fields, takes the form of a natural UV cutoff for the spectrum of
the canonical scalar differential operator on the manifold, the Laplacian:
the space of physical fields is assumed to be spanned by only those
eigenfunctions of the Laplacian whose eigenvalues are below the UV cutoff
value, which may be taken to be at the Planck scale, for example.

It has been shown that
fields on generic curved backgrounds, i.e., on Riemannian manifolds, which are bandlimited in this sense,
indeed can be precisely reconstructed everywhere, merely from the knowledge of
the field's amplitude samples on an arbitrary lattice - if the lattice' average
spacing is at the UV cutoff scale. The proof strategy is based on the principle underlying
sampling theory that we discussed above. Concretely, on generic compact Riemannian
manifolds, this covariant space of bandlimited functions possess a finite
dimension, $N$. By the general principle, this dimension, which depends on the volume
of the manifold, is also the number of
sample points required for the reconstruction of fields. The limit $N\rightarrow\infty$ was then studied and, under
mild assumptions, it was shown that, as the volume of the compact Riemannian
manifold is made to grow, the number of sample points required to perfectly
reconstruct a field asymptotically increases proportionally, \cite{ak-prl-2008}. This means that
a finite density of samples indeed suffices also in the large volume limit.

The sampling-theoretic natural UV cutoff for fields on curved space has been
applied to the relatively simple case of the space-like hypersurfaces in
inflationary cosmology. Possible signatures in the scalar and tensor spectra
of the cosmic microwave background have been calculated and discussed, see {\it
e.g.} \cite{cosmo}. In this work, the UV cutoff (i.e., the bandwidth) was
implemented on the individual
spacelike hypersurfaces through the generalized uncertainty principle.
Work on inflation with fields that obey a fully
spacetime-covariant sampling theoretic natural UV cutoff is in progress.

For sampling theory to become a useful mathematical tool, not only
for QFT on curved space, but also for quantum gravity, sampling theory
needs to be generalized so that it applies not only to fields on fixed backgrounds but
also to space and spacetime itself.
To this end, it has been shown in \cite{ak-prl-2009} that, under certain conditions and under the
assumption of the sampling-theoretic natural UV cutoff on the spectrum of the
Laplacian, also the shape of a curved space, i.e., of a Riemannian manifold, can
be reconstructed - up to structures that would be smaller than the UV cutoff
scale. Under the assumption of the sampling theoretic natural UV
cutoff, curved spaces are physically indistinguishable if they differ only on
scales smaller than the cutoff scale.

We will review this finding below and we
will here then derive the explicit reconstruction formulas for curved spaces and the fields on them. We will also take first steps towards the generalization of sampling
theory to the sampling and reconstruction of curved spacetimes, i.e.,
Lorentzian manifolds.

\section{Sampling theory of fields}

Let us now consider in detail the generalization of sampling theory to physical fields in curved
Euclidean-signature spacetimes, such as the fields that are being summed over
in the path integral of Euclidean quantum field theory (QFT),
\cite{ak-prl-2008,ak-prl-2004}.

To this end, consider a spacetime described by a compact smooth Riemannian
manifold. For simplicity, we assume that it has no boundary. For the covariant
inner product of fields on the manifold we use the usual bra-ket notation
inspired by first quantization: $(\phi\vert\psi)=\int d^dx~\sqrt{\vert
g\vert}~\phi(x)\psi(x)$, so that one has, for example: $\phi(x)= (x\vert\phi)$.
We use the sign convention in which the spectrum of the Laplacian is positive
and we choose $c=\hbar=G=1$. The Laplacian is self-adjoint, with $\Delta
v_{\lambda_i}=\lambda_i v_{\lambda_i}$. Since the Laplacian's inverse is compact, the eigenvalue problem is solved by normalizable eigenfunctions
with discrete eigenvalues that do not possess an accumulation point.
As already mentioned, the generalization of the assumption of
bandlimitation is the assumption that there exists a natural hard UV cutoff,
$\Lambda$, of the spectrum of the Laplacian, with $\Lambda$, for example, at
the Planck scale. In QFT, the space of fields, ${\cal F}$, that is being
integrated over in the path integral, is then spanned by the eigenfunctions
$v_{\lambda_i}$ of the Laplacian whose eigenvalues, $\lambda_i$, are below the
cutoff, $\lambda_i <\Lambda$. Let $P$ denote the projector onto ${\cal F}$ and
let us denote the Laplacian restricted to ${\cal F}$ by $\Delta_c = \Delta
\vert_{\cal F}$. The fields $\vert\phi)\in{\cal F}$ that occur in the path
integral obey $\phi(x)= (x\vert\phi)=(x\vert P\vert \phi)$. This
means that the point-localized fields $\vert x )$ are now indistinguishable
from the fields $P\vert x )$ in which wavelengths shorter than the cutoff scale
are removed. Intuitively, this expresses a minimum length uncertainty
principle.

How then does sampling theory arise? Since the Laplacian's spectrum does not
possess accumulation points, the
dimension, $N$, of the space of fields is finite, dim$({\cal F})=N$. By the general principle underlying sampling theory it is clear that, therefore,
any field $\phi(x)\in{\cal F}$ can be perfectly reconstructed
everywhere if known only on $N$ generic points of the manifold. The fields,
actions and equations of motion therefore possess a representation on the
smooth spacetime manifold as well as equivalently also on any lattice of $N$
generic points. Now if the volume, $V$, of the manifold is increased, then also the number, $N$, of eigenvalues below the cutoff increases. This is clear because as the manifold is extended to infinite size the spectrum of the Laplacian has to become continuous. The key question therefore is, how fast the dimension $N$ increases as a function of the volume. If it increases linearly we maintain a sampling theorem because then a finite ratio of $N/V$, i.e. a finite density of sampling points suffices to reconstruct the bandlimited fields on the manifold. Indeed, under mild conditions, Weyl's asymptotic formula
implies that as the infrared (IR) cutoff is removed by letting the volume of the
manifold diverge, $V\rightarrow\infty$, one has that $N\rightarrow \infty$ proportionally, i.e., such that
the density of samples necessary for reconstruction, $N/V$, indeed stays finite
\cite{ak-prl-2008}.

\section{Sampling theory of spacetimes} \rm

Assuming the natural hard UV cutoff above, is it possible to reconstruct also the very shape (curvature and
global topology) of a Euclidean-signature spacetime
from suitable samples taken at a discrete set of points?

In order to address this question, let us recall that, while it is not necessary, it is often convenient for the purpose of terminology and intuition to think of a Riemannian manifold as being embedded. This is because one may then visualize its curvature as a ``shape". Let us adopt this intuitive terminology for Riemannian and, in a looser sense, also for Lorentzian manifolds, while keeping in mind that the purpose of the terminology is only to aid intuition. We will not actually mathematically embed the manifolds.

In this terminology, what we call a spacetime's ``shape" is usually
best described in terms of the affine
connection and, or, the metric tensor. It is possible, however, to describe a manifold's shape also by different means.  Let us
recall a comment by Einstein \cite{Einstein}, who pointed out that the
nontrivial shape of a manifold manifests itself not only in the nontriviality
of the parallel transport of tensors. Crediting Helmholtz, Einstein emphasized
that the shape of a manifold can also be thought of in terms of the
nontriviality of the mutual distances among points: In $d$-dimensional flat
space, consider $M$ points. In cartesian coordinates, the points possess $M d$
coordinates $x^{(n)}_i$ with $n=1,...,M$ and $i=1,...,d$. By Pythagoras, the
$M(M-1)/2$ mutual distances $s_{n,n'}$ obey the equations $s_{n,n'}^2 =
\sum_{i=1}^d (x^{(n)}_i-x^{(n')}_i)^2$. If $M>2d+1$, the $M d$ coordinates can
be eliminated in these $M(M-1)/2$ equations, to leave $M(M-1)/2 -M d$
nontrivial equations that must hold among the mutual distances $s_{n,n'}$ if
the manifold is indeed flat. If the manifold is curved this manifests itself in
the way in which these equations are violated.

For example, let us consider a curved manifold, say of two dimensions, of some finite size. Choose some $N$ more or less evenly spread-out points on the manifold and record a table of their mutual geodesic distances (assuming, for simplicity, their uniqueness). Then, generically, knowledge of this table of distances alone should suffice to reconstruct a skeleton of the manifold. While that skeleton captures the shape of the manifold on large scales it of course leaves the shape undetermined in between the discrete points.

As we will discuss later, this approach should apply not only to Riemannian but also to Lorentzian manifolds. Given a table of invariant distances between $N$ events, a type of skeleton of the ``shape" of the Lorentzian manifold would be determined. As we will discuss later, this approach could open up new methods for the spectral geometry of Lorentzian manifolds. Recall that Lorentzian spectral geometry is  notoriously hard because of the hyperbolic nature of the d'Alembertian as opposed to the elliptic Laplacian.

For now, let us continue with the case of Riemannian manifolds. It is clear that, in a quantum theory, the measurement of geodesic distances may not be practical nor unique nor well-defined. Let us therefore consider replacing the notion of geodesic distance with a different notion of distance that is adapted to quantum field theories. Namely, let us measure the ``distance" between points through the amplitude of the correlator, i.e., of the two-point function. Intuitively, the correlator of two points is a measure of distance because, quantum field theoretically, it is a measure of vacuum entanglement, which drops with distance. The same conclusion is reached, of course, when the correlator is interpreted in terms of the heat equation.

We will now investigate how to reconstruct the shape of a euclidean-signature spacetime of finite volume
by sampling at a sufficient number, $N$, of generic points  the
propagator, or correlator, $G(x^{(n)},x^{(n')})=(x^{(n)} \vert P (\Delta +m^2)^{-1}
P\vert x^{(n')} )$ of a scalar field for each
pair of the $N$ chosen points. Generally, the larger the distance between
$x^{(n)}$ and $x^{(n')}$, the smaller is the correlator.  Indeed, with caveats that we will discuss below, the knowledge of the $N(N-1)/2$ matrix elements
$G(x^{(n)},x^{(n')})$ suffices to reconstruct the shape of the spacetime up to
the UV cutoff scale. To see this, we note first that the matrix
$(G(x^{(n)},x^{(n')}))_{nn'}$ represents the correlator, $(\Delta_c
+m^2)^{-1}$, in a basis, namely the basis $\{P\vert x^{(n)} )\}$, which means
that we can determine its eigenvalues, as we will explicitly work out in the following section. Since the correlator is diagonal in the
same basis as the Laplacian $\Delta_c$, we also obtain the spectrum of
$\Delta_c$.
(Also in interacting theories the Laplacian's spectrum can be
calculated since the functional dependence of the correlator on $\Delta_c$ is
still determined by the action.) Crucially now, the eigenvalues of $\Delta_c$ largely determine the shape of the spacetime, from large length scales down to the cutoff scale.

To see this, let
us recall key results of the discipline of spectral geometry. Spectral geometry investigates the
relationship between the shape of a manifold (or its domain) and the spectrum of
its Laplacian or Dirac operator. Notice that spectral geometry thereby naturally
combines functional analysis and differential geometry, {\it i.e.}, the
mathematical languages of quantum theory and general relativity.
 It is a simple fact, of course, that isometric manifolds are isospectral, i.e., that their Laplacians possess identical spectra.
The discipline of (``inverse") spectral geometry is concerned with the nontrivial converse, namely with the study of the extent to which the spectra of manifolds determines their metrics. It is believed that the spectra largely determine the shape of a manifold, except for special circumstances. Indeed, special counter examples are known. Applied to the case of vibrating membranes, for example, this means that ``one cannot always hear the shape of a drum", see, {\it e.g.},
\cite{Gordon}. Even pairs of isospectral but non-isometric manifolds that are
compact and simply connected have been constructed \cite{Schueth}. It is known,
however, that the eigenvalues do indeed change continuously as a function of the shape of
the manifold. Also, the eigenvalues are nondegenerate for generic manifolds,
and a manifold can have degenerate eigenvalues only if it possesses a
continuous group of isometries \cite{Bando}. Below, we will discuss the question of the uniqueness of inverse spectral geometry further.

For our purposes, since we assume an ultraviolet cutoff,
we are led to consider classes of manifolds whose Laplacians
share the same eigenvalues (and their multiplicities) only up to the cutoff
$\Lambda$, and which we may therefore call $\Lambda$-isospectral. Indeed,
the samples of the matrix elements of the correlator,
$(G(x^{(n)},x^{(n')}))_{nn'}$, determine only the $N$ eigenvalues of the
Laplacian $\Delta_c$. The eigenvalues that the full Laplacian, $\Delta$,
possesses beyond the cutoff remain undetermined. This shows us what the UV
cutoff means for the shape of spacetime itself. The cutoff does not directly
imply a cutoff for the curvature, for example. Instead, the fact that the
eigenvalues of the Laplacian beyond the cutoff remain undetermined by any
measurement possible means that all $\Lambda$-isospectral manifolds are
physically indistinguishable and are therefore to be placed into one equivalence class.

In order to adopt an unambiguous terminology, let us use the term
``spacetime with UV cutoff", specified by $N$ eigenvalues, $\lambda_1 \le
...\le \lambda_N<\Lambda$, to describe the corresponding equivalence
class of $\Lambda$-isospectral manifolds.

In this context, let us note an interesting detail.
First, consider functions on a fixed background manifold. For these functions, wavelength and amplitude are separate properties. For example, there are functions that describe waves that have short wavelength but large amplitude. The situation is different for wrinkles of curvature in the underlying space itself. For wrinkles in a Riemannian manifold, wavelength and amplitude are not separate properties. If the ``amplitude" of a short-wavelength wrinkle is made to grow, that wrinkle acquires a long wave length.

Now in spectral geometry, the higher the eigenvalues of the Laplacian the higher the
``squared momentum" that they represent, and, intuitively, therefore the smaller the wrinkles which they determine in the manifold. We expect that
the eigenvalues up to $\Lambda$ can only
determine the shape of a spacetime with UV-cutoff from large scales
down to lengths as small as the cutoff scale. The undetermined eigenvalues
beyond the cutoff would describe wrinkles on length scales smaller than the
cutoff scale. At distances smaller than the cutoff scale the shape of a
spacetime with UV cutoff is not determined.

This can be viewed in terms of representation theory. In general relativity,
the choice of coordinate system is merely a choice of representation for an
underlying Riemannian manifold. With the UV cutoff, even the choice of
Riemannian manifold is merely a choice of representation for an underlying
``spacetime with UV cutoff" that is defined through the first $N$ eigenvalues
of the Laplacian.

The complete picture is more subtle, however. A theory may
contain additional fields with interactions that allow one to physically
distinguish among certain $\Lambda$-isospectral manifolds. At the very least,
we will have to divide each equivalence class of $\Lambda$-isospectral manifolds
into sub-equivalence classes of manifolds that are continuously deformable into
each other within their class of $\Lambda$-isospectral manifolds. This is
because at least for those subclasses, and possibly also for sub-subclasses
within them, we can define what we may call ``geometric quantum numbers" that
distinguish them and that could be measurable in the full theory.

For example, parity and the dimension of the manifold can be viewed as a geometric quantum number. The dimension is 
measurable, \it e.g., \rm through interactions involving tensors (whose
dimensions indicate the manifold's dimension). Indeed, $\Lambda$-isospectral
manifolds of different dimensions cannot be continuously deformed into another
\cite{Bando}. Note that the scaling of the Laplacian's spectrum for
asymptotically large eigenvalues is in one-to-one correspondence to the
manifold's dimension \cite{piiss}. It should be interesting to study the set of
possible geometric quantum numbers, and their relation to cohomology, by
methods similar to those used to construct isospectral non-isometric manifolds,
see {\it e.g.} \cite{Schueth}.

A ``spacetime with UV cutoff" is, therefore, an equivalence class of manifolds
that are $\Lambda$-isospectral and possess the same geometric quantum numbers.
An intriguing possibility is that it may not be necessary to keep track of
geometric quantum numbers as variables that are separate from the spectrum
after all, namely when working with the Laplacians on general tensors, including the Laplacian $d\delta+\delta d$ on
all differential forms, and the Dirac operator. Their spectra may well include
all information about the geometric quantum numbers. We will further develop this issue in Sec.8.

\section{Explicit sampling of fields and manifolds}

Let us now develop the sampling theory for both spacetime and field, for the case of the Laplacian on 0-forms. The generalization to Laplacians on all tensors and the Dirac operator should be possible with the same strategy. The procedure was first outlined in \cite{ak-prl-2009}. Here, we explicitly carry out this program.

In our framework, a (compact, euclidean-signature) spacetime with ultraviolet cutoff is specified by
the eigenvalues $\lambda_1,...,\lambda_N$ of $\Delta_c$ (and possibly by geometric quantum numbers and/or by the spectra of further differential operators). A field, $\vert \phi
)$, on the spacetime is a vector in the $N$-dimensional Hilbert space on which
$\Delta_c$ acts, conveniently specified through its coefficients $\phi_i$ in an
ON eigenbasis $\{\vert v_{\lambda_i} )\}$ of $\Delta_c$, as $\vert\phi )=
\sum_{i=1}^N \phi_i \vert v_{\lambda_i} )$. Now consider a continuous
representation of the spacetime as a manifold that possesses the specified
geometric quantum numbers, such as the dimension, and whose Laplacian on 0-forms, $\Delta$,
possesses the spectrum of $\Delta_c$ up to $\Lambda$. After we choose coordinates, the abstract eigenvectors of $\Delta_c$ are then represented as eigenfunctions $v_{\lambda_i}(x)$
of $\Delta$. For example, the field $\phi$ is represented by the function
$\phi(x)=\sum_{i=1}^N\phi_i v_{\lambda_i}(x)$.

Because of the presence of the UV-cutoff, both the continuous field and the underlying spacetime possess completely equivalent discrete representations as well.    We obtain such a lattice
representation by choosing any $N$ generic points $x^{(1)},...,x^{(N)}$ at which we
sample and record the matrix of correlators $G(x^{(i)},x^{(j)})$ and the
field's amplitudes $\phi(x^{(i)})$. Crucially then, a sampling theorem holds: from these data sampled at discrete points one can fully reconstruct
the $\lambda_1,...,\lambda_N$ and $\phi_1,...,\phi_N$ which, as we saw, specify
the abstract spacetime and field, which in turn possess continuous representations as Riemannian manifolds and continuous functions on them.

Concretely, assume that the matrix $\tilde{C}$ of correlators for the cut-off Laplacian $\Delta_c=P\Delta P$ for all pairs of the discrete sample points $x^{(i)}$ is explicitly known:
\begin{equation}
\tilde{C}_{ij} := (x^{(i)}\vert P \frac{1}{\Delta + m^2}P\vert x^{(j)})
\end{equation}
Recall that the projector $P$ onto the subspace of physical fields, i.e., onto the space of bandlimited fields, is given by $P= \sum_{i=1}^N \vert v_{\lambda_i})(v_{\lambda_i}\vert$. Thus:
\begin{eqnarray}
\tilde{C}_{ij} &=& \sum_{a,b=1}^N(x^{(i)}\vert v_{\lambda_a})(v_{\lambda_a}\vert \frac{1}{\Delta + m^2}\vert v_{\lambda_b})(v_{\lambda_b}\vert x^{(j)})\\
   &=& E_{ia}C_{ab}E^\dagger_{bj}
\end{eqnarray}
In the last line, we defined an as yet unknown matrix $E$ with the matrix elements $E_{ia}:= (x^{(i)}\vert v_{\lambda_a})$, and an as yet unknown diagonal matrix $C$ with the matrix elements $C_{ab}:= \lambda_a\delta_{ab}$. We adopted the Einstein summation convention. What we are after first is the reconstruction of the spectrum of $\Delta_c$, i.e., of the eigenvalues of $C$. First, we find:
\begin{equation}
C= E^{-1}\tilde{C}{E^\dagger}^{-1}\label{dfr}
\end{equation}
Also, for the given discrete set of points $\{x^{(i)}\}$, let us assume explicitly given the inner products between the vectors $\vert x^{(i)})$ which represent these points. Because of the ultraviolet cutoff, the matrix of these inner products, $\tilde{B}$, has the matrix elements:
\begin{eqnarray}
\tilde{B}_{ij} &:=& (x^{(i)}\vert P\vert x^{(j)})\\
 &=& \sum_{r=1}^N (x^{(i)}\vert v_{\lambda_r})(v_{\lambda_r}\vert x^{(j)})\\
 &=& E_{ir}E^\dagger_{rj}
 \end{eqnarray}
We therefore have $E^{-1}\tilde{B}=E^\dagger$, and thus:
\begin{equation}
{E^\dagger}^{-1}={\tilde{B}}^{-1}E
\end{equation}
With Eq.\ref{dfr}, this finally yields:
\begin{equation}
C = E^{-1}\tilde{C}\tilde{B}^{-1}E \label{hg}
\end{equation}
Recall that we assume that the matrices $\tilde{B}$ and $\tilde{C}$ are explicitly known, from the lattice formulation of the theory. In Eq.\ref{hg}, due to the adjoint action of $E$, the characteristic polynomials
of $C$ and $\tilde{C}\tilde{B}^{-1}$ agree. This means that by diagonalizing the known matrix $\tilde{C}\tilde{B}^{-1}$ we obtain on one hand the matrix $E$, with $E_{nj}=( x^{(n)}\vert v_{\lambda_j} )$, of the diagonalizing change of basis. On the other hand, we obtain
the spectrum of the diagonalization, $C$. The spectrum of $C$ then of course easily yields the spectrum of $\Delta_c$, because both are diagonal in the same basis.

Let us note that, interestingly, if there are two or more scalar fields in the theory, the spectrum of $\Delta_c$ can be reconstructed from lattice data by a second method which does not require knowledge of the matrix $B$ of position overlaps: if there are two scalar fields of different masses $m$ and $M$, for example, we can assume known for a set of $N$ discrete points $\{x^{(i)}\}_{i=1}^N$ the correlator matrix $\tilde{C}$ and also the correlator matrix $\tilde{D}$ with matrix elements:
\begin{equation}
\tilde{D}_{ij} := (x^{(i)}\vert P \frac{1}{\Delta + M^2}P\vert x^{(j)})
\end{equation}
We have
\begin{equation}
\tilde{C} = ECE^\dagger, ~~~~~\tilde{D} = EDE^\dagger
\end{equation}
Here, $C$ and $D$ are the matrices of the diagonalizations of $1/(\Delta_c+m^2)$ and $1/(\Delta_c+M^2)$ respectively, as before. We have  $C=E^{-1}\tilde{C}{E^\dagger}^{-1}$ and $D=E^{-1}\tilde{D}{E^\dagger}^{-1}$, and therefore ${\tilde{D}}^{-1}ED={E^\dagger}^{-1}$. Thus:
\begin{equation}
C=E^{-1}\tilde{C}\tilde{D}^{-1}ED
\end{equation}
Since $D^{-1}=C^{-1}+M^2-m^2$ this yields $C=E^{-1}\tilde{C}\tilde{D}^{-1}E(C^{-1}+M^2-m^2)^{-1}$ and therefore, finally:
\begin{equation}
(1+(M^2-m^2)C) = E^{-1}\tilde{C}\tilde{D}^{-1}E \label{ed}
\end{equation}
Eq.\ref{ed} shows that by diagonalizing the known matrix $\tilde{C}\tilde{D}^{-1}$, we can obtain the diagonalizing matrix $E$, the spectrum of $C$ and therefore the sought-after spectrum of $\Delta_c$. Note that we recover the first reconstruction method for the spectrum of $\Delta_c$, which used only one scalar field, by letting $M\rightarrow\infty$. This is because in this limit $\tilde{D}\rightarrow\tilde{B}$.

Now in order to reconstruct also the continuous field $\Phi$, we need to calculate the coefficients
$\phi_j=( v_{\lambda_j}\vert\phi )$. To this end, we insert a resolution of the identity in
$( x^{(n)} \vert \phi  ) = \sum_{j=1}^N ( x^{(n)}\vert v_{\lambda_j} ) ~(
v_{\lambda_j} \vert \phi  )$, {\it i.e.}, $\phi(x^{(n)})=\sum_j E_{nj}~\phi_j$.
Since the
$\phi(x^{(i)})$ are known samples, we obtain $\phi_i=\sum_j
(E_{ij})^{-1}~\phi(x^{(j)})$. Thus, from the samples of field amplitudes and
correlators, we have obtained the spacetime in terms of the eigenvalues of its
Laplacian $\Delta_c$ and the field as a vector in the $N$-dimensional vector
space on which $\Delta_c$ acts. At this point we are free to represent the
abstract spacetime by the same or any other member of its equivalence class of
$\Lambda$-isospectral manifolds, choose coordinates and express the field as an
explicit function. The choice of real-valued orthonormalized
eigenfunctions, while normally ambiguous up to a factor of $-1$, is here fixed
by continuity. The formalism establishes, therefore, an equivalence between
discrete and continuous representations of spacetimes and fields.

\section{Extending spectral geometry}

Our use of spectral geometry so far has been limited to the use of the Laplacian on 0-forms. Let us now investigate under which circumstances, for the purpose of inverse spectral geometry and sampling theory, the spectrum of the Laplacian on 0-forms should be augmented by the spectra of other differential operators, such as the Dirac operator or Laplacians on tensors. This will yield a possible strategy for developing inverse spectral geometry further. The key idea is to consider the spectral geometry of infinitesimal changes to both the spectrum and the shape of a manifold. The advantage of considering infinitesimally small changes to the shape of a manifold lies in the fact that these small shape changes can be parameterized in terms of functions on the manifold. This means that the shape changes can be  described within the function space of the manifold and therefore within the framework of the sampling theory of functions, rather than the more difficult (because nonlinear) sampling theory of manifolds.

To this end, consider an arbitrary compact Riemannian manifold, $(M,g)$, and the spectrum, $\{\lambda_i\}_{i=0}^\infty$ of its Laplacian on 0-forms. If we infinitesimally perturb the manifold, this entails an infinitesimal change to its spectrum. In the simplest case, we may describe a small deformation of the manifold by specifying a scalar field, $\varphi$, on the manifold, in the same way that the deformation of the membrane of a drum, when played, can be specified by a scalar amplitude field, $\phi$. The scalar field $\phi$ that describes the deformation of the manifold can then be expanded canonically in terms of the eigenfunctions of the Laplacian on 0-forms. This means that the small deformation of the manifold is specified by a sequence of small scalar numbers $\{\varphi_i\}_{i=1}^\infty$, namely the coefficients of $\varphi$ in the Laplacian's eigenbasis. The corresponding small change in the spectrum of the Laplacian on 0-forms is described by a sequence of small numbers $\{\Delta \lambda_i\}_{i=0}^\infty$. Clearly, the sequence $\{\varphi_i\}_{i=1}^\infty$ determines the sequence $\{\Delta \lambda_i\}_{i=0}^\infty$, because the metric determines the spectrum. But does $\{\Delta \lambda_i\}_{i=0}^\infty$ determine the sequence $\{\varphi_i\}_{i=1}^\infty$?

To address this question, it will be useful to implement the UV-cutoff. This is because, in this case, the two sequences $\{\Delta \lambda_i\}_{i=0}^\infty$ and  $\{\varphi_i\}_{i=1}^\infty$ are truncated to only the first $N$ elements. The question is then translated into the question of whether the map, $\tau$, which maps $\{\varphi_i\}_{i=1}^N$ to $\{\Delta \lambda_i\}_{i=0}^N$ is invertible (at the origin). Indeed, generically, we may expect $\tau$ to be invertible, as we have a map from $R^N$ into $R^N$ and the determinant of the Jacobian has no obvious reason to vanish.

An important fact, however, is that only the simplest deformations of a Riemannian manifold can be described through a scalar function. For example, in cosmology, small deformations of spacelike hypersurfaces of spacetime, away from flatness, are described in terms of scalar, vector as well as tensor fluctuations of the metric, i.e., in terms of scalar, vector and tensor fields. We can expand these fields in terms of eigenfunctions of the Laplacians on forms and on tensors.

 The fact that the description of a small deformation of a generic Riemannian manifold requires the specification of not only a scalar function but also for example of vector and tensor functions is important, as becomes clear once we consider the case with a UV-cutoff.
This is because, with the UV-cutoff, a small deformation of a Riemannian manifold can require the specification of not just the sequence $S_s$ of coefficients  $\{\varphi_i\}_{i=1}^N$ of a scalar field, but also for example the sequences $S_v$ and $S_t$ of coefficients (in Laplacian eigenbases) of vector and tensor fields. The combined dimension of $S_s, S_v$ and $S_t$ is clearly much larger than $N$. This means that the map $\tau$ that maps the small shape deformation into the corresponding small perturbation of the spectrum, $\{\Delta \lambda_i\}_{i=0}^N$, cannot be invertible. Indeed we now obtain in this way a handle on how many additional spectra of further differential operators such as the Laplacian on forms and general tensors and the Dirac operator are needed to enable generic invertibility of the map $\tau$ that maps the metric into the spectra of differential operators. Work in this direction is in progress, \cite{ak-eb}.

Let us finally also briefly address the question of how this framework of sampling theory can be extended to Lorentzian manifolds and the fields on them. Indeed, spectral geometry has been mostly confined to Riemannian manifolds so far because of the fundamental fact that the d'Alembert operator is hyperbolic in nature, unlike the Laplacian which is elliptic. As a consequence, it is nontrivial to suitably discretize the spectrum of the d'Alembertian. But the discretization of the spectrum  is of course necessary to start a program of spectral geometry in which a discrete spectrum carries the information about the shape of a manifold. In this context, let us recall, however, Helmholtz' argument and our implementation of it above: a table of the mutual distances between $N$ points reflects the flatness or curvature of the underlying space. Intuitively, the table of distances allows one to build a skeleton of the manifold that approximates its shape. Above, we used this idea for the case of Riemannian manifolds by using the amplitudes of correlators as proxies for the distances between points. In the case of Lorentzian manifolds, one may sample instead the correlator or propagator between pairs of $N$ events. Even though the correlator or propagator now provides a proxy for invariant positive and negative proper squared distances, these too should allow one to build a kind of skeleton of the manifold that approximates what might be called its shape, i.e., its spacetime curvature. Interestingly, by this method we arrive, as above in the Riemannian case, at a $N \times N$ matrix for the d'Alembertian. This means that we arrive at a natural cutoff and discretization of the d'Alembertian's spectrum. It should be interesting to investigate to what extent this discretized d'Alembertian can serve as a basis for both, the inverse spectral geometry and the sampling theory for Lorentzian manifolds and the fields in them.

\section{Summary and Outlook}

We showed that spacetime could be simultaneously continuous and discrete, in the same way that information can. To this end, we considered the cutting off of the spectrum of the Laplacian (or d'Alembertian) at an eigenvalue close to the Planck scale. We found that, in this case, physical theories possess equivalent continuous and discrete representations, and that external symmetries can be fully preserved. While we made some progress towards adapting sampling theory and spectral geometry to Lorentzian manifolds, clearly much more work in this direction is required. Further, it should be very interesting to go beyond the study of the kinematics and to investigate the dynamics. First steps were taken in \cite{ak-prl-2009}, by considering the number, $N$,  of sample points needed for reconstruction of fields on a compact Riemannian manifold with cutoff, and the reconstruction of the manifold itself.  $N$ is the number of eigenvalues of the Laplacian, in the simplest case of the Laplacian on scalar functions, which are below the cutoff. Interestingly, this number, $N$, has an expansion in terms of the curvature: $N=\int d^4x ~\sqrt{|g|}~(c_0 + c_1 R + O(R^2))$. Up to corrections that are of higher order in the Planck scale, this is of the form of the euclidean Einstein action, i.e., $S= (16\pi c_1)^{-1}~N =  (16\pi c_1)^{-1} ~Tr(1)$. Here, before renormalization, $c_0$, is related to the (unrenormalized) cosmological constant $\Lambda$ via $\Lambda=c_0/16\pi c_1$. When there is no curvature, $c_0$ expresses the density of degrees of freedom in the theory: $c_0 = N/V$ with $V=\int d^4x~\sqrt{|g|}$. Curvature then is acquiring a new interpretation. In addition to expressing the nontriviality of parallel transport, curvature is here seen to be the local modulation of the density of degrees of freedom in the theory. It should therefore be possible to re-express also the variational principle as an extremization with respect to the density of degrees of freedom, or their overall number in a finite volume. In effect, we are diagonalizing the Einstein action by expressing it in the Laplacian's eigenbasis. A key challenge will be to express and investigate also a bandlimited formulation of interacting theories of the standard model in this basis. This will introduce the Dirac operator and generalized Laplacians on tensors, which nicely fits with our observation in Sec.8 that the spectra of such operators are in any case needed to complete the information theoretic description of spacetimes.   
\bigskip\newline
\bf Acknowledgement. \rm This work has been supported by the Discovery and Canada Research Chair Programs of National Science and Engineering Research Council (NSERC) of Canada.


\section*{References}
\begin{thebibliography}{10}

\bibitem{discreteqg} C. Rovelli, \it Quantum Gravity, \rm CUP, Cambridge (2004)

\bibitem{Goedel} K. G\"odel, \it The Consistency of the Continuum-Hypothesis, \rm Princeton
University Press (1940)

\bibitem{Cohen} P.J. Cohen, Proc.
Natl. Acad. Sciences {\bf 50} (6): 1143�1148 (1963), P.J. Cohen, \rm Proc. Natl. Acad. Sciences
{\bf 51} (1): 105�110 (1963)

\bibitem{Isham} Isham C.J., Butterfield J. 2000, \rm Found. of Physics, {\bf 30}, 10, p.
1707-1735

\bibitem{ak-prl-2009} A. Kempf, Phys. Rev. Lett. {\bf 103}, 231301 (2009)

\bibitem{shannon} C. E. Shannon, W. Weaver \it
The Mathematical Theory of Communication, \rm Univ. of Illinois Press (1963),
J.J. Benedetto, P.J.S.G. Ferreira, \it Modern Sampling Theory, \rm Springer
Verlag, Heidelberg (2001), P.J.S.G. Ferreira, A. Kempf, IEEE Trans. Sign. Proc.
\bf 54, \rm 3732 (2006)

\bibitem{ak-prl-2000} A. Kempf, Phys. Rev. Lett. {\bf 85}, 2873 (2000)

\bibitem{ucrs} D.J. Gross, P.F. Mende, Nucl. Phys. {\bf B303}, 407 (1988),~~
D. Amati, M. Ciafaloni, G. Veneziano, Phys. Lett. {\bf B216} 41, (1989),  D.V. Ahluwalia, Phys. Lett.
{\bf B339}, 301 (1994), M.-J. Jaeckel, S. Reynaud, Phys. Lett. {\bf A185}, 143
(1994), E. Witten, Phys. Today {\bf 49} (4), 24 (1996), G. Amelino-Camelia, J.
Ellis, N.E. Mavromatos, D.V. Nanopoulos, Mod. Phys. Lett.{\bf A12} 2029 (1997)

\bibitem{ak-qheisenberg} A. Kempf, J. Math. Phys. {\bf 35} (9), 4483 (1994)

\bibitem{Gambini} R. Gambini, J. Pullin, Phys. Rev. {\bf D59}, 124021 (1999)

\bibitem{cosmo} A. Kempf,  Phys. Rev. {\bf D63} 083514, (2001),
 A. Kempf, J. C. Niemeyer, Phys. Rev. {\bf D64} 103501 2001),
R. Easther, B. R. Greene, W. H. Kinney, G. Shiu, Phys. Rev. {\bf D64} 103502
(2001), R. Easther, W.H. Kinney, H. Peiris, JCAP \bf 05 \rm 009 (2005)

\bibitem{ak-lattice} A. Kempf,  Europhys. Lett. {\bf 40}, 257 (1997)

\bibitem{Amelino} G. Amelino-Camelia, Int. J. Mod. Phys. {\bf D11}, 35 (2002)


\bibitem{ak-prl-2008} A. Kempf, R. T. Martin, Phys. Rev. Lett. {\bf 100} 021304 (2008)

\bibitem{ak-prl-2004} A. Kempf, Phys. Rev. Lett. {\bf 92}, 221301 (2004), A.
Kempf, Phys. Rev. \bf D69, \rm 124014 (2004)

\bibitem{Einstein} A. Einstein, \it Grundz\"uge der Relativit\"atstheorie, \rm
p12,  Vieweg \& Sohn, Baunschweig, 5th Edition (1979)


\bibitem{Gordon} C. Gordon, D.L. Webb, S. Wolpert,
Bull. Amer. Math. Soc. {\bf 27}, 134 (1992)

\bibitem{Schueth} D. Schueth, Ann. of Mathematics, {\bf 149}, 287 (1999)

\bibitem{Bando} S. Bando, H. Urakawa, T\"uhoku Math. J. \bf 35, \rm 155 (1983)

\bibitem{piiss} S. I. Andersson, M.L. Lapidus (Eds) \it Progress in Inverse
Spectral Geometry, \rm p.5, Birkh\"auser, Boston (1997)

\bibitem{ak-eb} A. Kempf, E. Brandao, in preparation


\end{thebibliography}
\end{document}